\documentclass[reprint,aps,pre,amssymb]{revtex4-1}

\usepackage[T1]{fontenc}
\usepackage{ucs}
\usepackage[utf8x]{inputenc}

\usepackage[reqno,intlimits]{amsmath}
\usepackage[final]{graphicx}
\usepackage{float}

\usepackage{color}

\renewcommand{\d}{\mathrm{d}}

\newcommand{\pb}[1]{
  \parbox[0pt][#1][c]{0cm}{}
}

\newcommand{\ew}[1]{
  \left\langle #1 \right\rangle
}

\begin{document}
  \allowdisplaybreaks
  \DeclareGraphicsExtensions{.eps}

  \title{Canonical fitness model for simple scale-free graphs}
  \date{\today}
  \author{F. \surname{Flegel} and I. M. \surname{Sokolov}}
  \affiliation{Institut f\"{u}r Physik, Humboldt-Universit\"{a}t zu Berlin, D-12489 Berlin, Germany}

  \begin{abstract}
    We consider a fitness model assumed to generate simple graphs with power-law heavy-tailed 
    degree sequence: $P(k) \propto k^{-1-\alpha}$ with $0 < \alpha < 1$,
    in which the corresponding
    distributions do not posses a mean. We discuss the situations in which the model is used to produce a multigraph
    and examine what happens if the multiple edges are merged to a single one and thus a simple graph is built.
    We give the relation between the
    (normalized) fitness parameter $r$ and the expected degree $\nu$ of a node 
    and show analytically that it possesses non-trivial intermediate and final asymptotic behaviors.
    We show that the model produces $P(k) \propto k^{-2}$ for large values of $k$ independent of $\alpha$.
    Our analytical findings are confirmed by numerical simulations.
  \end{abstract}

  \maketitle

\section{Introduction}
Large systems of objects are often described in terms of networks, i.e.
are represented by abstract graphs composed of vertices (objects) and edges
(links, interconnections) \cite{Newman2001, AlbertJeong1999, Williams2002, Faloutsos1999, Govindan2000, Koch1999}.
The number of vertices is called the order and the number of edges the size of the graph \cite{BollobasModern}.
The number $k$ of incident edges of a given vertex is called its degree, and the probability distribution $P(k)$ the
degree distribution of the graph.
A special feature of many real world networks is that the probability $P(k) \propto k^{-\gamma}$
with some $\gamma > 1$, i.e. these graphs show scale-free nature.

The two classical models of Gilbert \cite{Gilbert1959} and Erd\"os-R\'enyi \cite{erdos1960} do not show this property.
The first one is also known
by the term \emph{Bernoulli graph}
\cite{Gilbert1959, Solomonoff1951}.
In this model which is usually denoted by $G_{M,p}$, one starts with a set of $M$ labeled nodes.
Each possible pair of vertices is linked with an edge with probability $p$. This leads to a simple graph
with no one-loops.
The other random graph model was introduced by
Erd\"os and R\'enyi \cite{erdos1960} as a statistical ensemble $G_{M,N}$ that contains all possible graphs with
$M$ nodes and $N$ links (without multiple edges and one-loops),
and whose members are all of equal statistical weight. Austin et al. \cite{austin1959} however suggested that
multiple edges should also be permitted.
These two classical models of Gilbert \cite{Gilbert1959} and Austin et al. \cite{austin1959} correspond
to canonical and grand canonical ensembles, respectively \cite{DoroComplex}, with respect to the number of edges. 
When referring to the
classical canonical model, we mean that
both multiple edges and one-loops are allowed (see p. 9 of \cite{DoroComplex}), i.e. it generates a multigraph.
The difference to the grand canonical model is that the number $N$ of links (i.e. the network's size) is fixed and is the
same for every realization of $G_{M,N}$.

In both classical models the degree distribution $P(k)$ tends
to a Poissonian $P(k) \simeq e^{-\lambda}\lambda^k/k!$ in the limit $N \rightarrow \infty$,
where $\lambda = \langle k \rangle$ denotes the mathematical expectation of the vertex's degree.
Therefore these models cannot mimic the scale-free behavior of many real-world networks.

The scale-free behavior can be reproduced by considering growing networks,
for example it can be generated by preferential attachments algorithms \cite{Barabasi1999}, or
by modifying classical approaches which leads to fitness models.

Thus, Caldarelli et al. \cite{CaldarelliPaper} suggested a fitness model where each vertex is assigned a
fitness parameter chosen from a distribution $\rho_f(f)$.
For each possible connection between two vertices the
probability that the connection exists is given by a linking function of the two fitnesses.
This corresponds to a generalization of Gilbert's model.
In the present work, we examine a certain graph generation model equivalent to the model suggested
by Goh et al. \cite{Goh2001} which is a generalization
of Austin et al's model.
The model can also be considered as a special case of Caldarelli et al. \cite{CaldarelliPaper}. 
The authors of \cite{Goh2001} assumed it as a trivial statement that the degree distribution follows
the fitness distribution in its asymptotics.
As we proceed to show, the approach of \cite{Goh2001} does not work for fitness distributions lacking the mean.
In what follows we discuss this situation
in detail and show that for heavy-tailed fitness distributions $\rho_f(f) \propto f^{-1-\alpha}$ with $0<\alpha<1$,
the degree distributions $P(k) \propto k^{-2}$ are universally produced independent of $\alpha$.
Thus our model provides another example of the inverse square laws observed in many 
fields of network research \cite{DeLosRios2001, Caldarelli2004, Goh2002}.
The model is therefore not applicable to mimic such networks as the network
of word co-occurrence ($\alpha = 0.8$, \cite{Cancho2001, Guillaume2006}) 
and the network of
homosexual contacts ($\alpha = 0.6$, \cite{Schneeberger2004}).

\section{Fitness Models}
As we pointed out above, both classical models can be altered in such a way that each vertex $i$
is assigned a positive random value $f_i$
which influences its chance to attract links in the building process. These values are called
the (intrinsic) \emph{fitness parameters}. For each vertex its fitness parameter is chosen
randomly according to a given probability density function (pdf) $\rho_f(f)$. Thus, the one-step procedure
of building the classical models is extended to a two-step procedure in which both fitness parameters
and degrees are determined randomly.

For the grand canonical model this generalization was investigated in \cite{CaldarelliPaper}:
For every pair $ij$ a link is drawn with a linking probability $F(f_i,f_j)$ where $F$ is a symmetric function in its arguments.
A natural choice for
the linking probability is $F(f_i,f_j) = (f_i f_j) / f_M^2$
where $f_M$ is the largest value of the $f$'s in the considered network.
Usually the distribution $\rho_f(f)$ relates to the intended degree distribution $P(k)$.

In our case we want to generate networks with degree distribution
$P(k) \propto k^{-1-\alpha}$ with $0<\alpha<1$, i.e. the ones lacking the mean.
We attempt to achieve this
by choosing $\rho_f(f) \propto f^{-1-\alpha}$.
The problem arises when we want to target a given number of edges because 
the expected number of edges $\frac{1}{2}\sum_{ij} f_i f_j /f_M^2$ is a strongly fluctuating quantity,
which is not the case for $\alpha > 1$. 
Hence, in the absence of the mean degree, this model provides only small control
over the networks size which urges us to concentrate on the canonical
fitness model.

\paragraph*{The Canonical Fitness Model}
starts again with $M$ vertices that have each an intrinsic fitness parameter distributed according to
the pdf $\rho_f$. We choose the probability $F(f_i, f_j)$  that a given link is connected to the nodes $i$ and $j$ to be
\begin{align*}
  F(f_i,f_j) &= \frac{f_i f_j}{\left(\sum_k f_k\right)^2}.
\end{align*}
At first, we will generate multigraphs and explain later how this model can be modified to produce simple graphs.

It can be easily seen that for each vertex $i$ with fitness parameter $f_i$ the degree expectation value
$\langle k_i \rangle = \lambda_i = 2Nf_i/\sum_j f_j$. This justifies the introduction of the normalized fitness parameters $r_i$:
\begin{align*}
   r_i = \frac{f_i}{\sum_{j=1}^M f_j} =  \frac{1}{1 + \frac{\sum_{j \neq i} f_j}{f_i}} \in (0,1),
\end{align*}
so that $\lambda_i = 2Nr_i$.
Thus, the distribution of degree expectations $\rho_\lambda(\lambda) = \rho_r(\lambda/2N)/2N$
has the same shape as the distribution of normalized fitnesses $\rho_r(r)$.
For $\alpha > 1$ the distribution of $r_i$ and therefore of $\lambda_i$ follows that of $f_i$,
but in the case of heavy-tailed distributions $\alpha < 1$ it differs in shape from $\rho_f(f)$ \cite{Eliazar2010}.
In what follows we concentrate on the behavior of $\rho_r(r)$ which gives the whole relevant information on the degree distribution.

Note, that even for narrow distributions of fitnesses the degree distribution differs from the fitness distribution.
It becomes obvious when considering the case $\rho _f (f)\propto \delta (f-f^\star )$ with some positive
value $f^\star $. Here both models are reduced to their respective classical model with its Poisson-shaped degree distribution.
Thus, we understand that for each vertex $i$ with the degree expectation $\lambda _i$, the probability $P_i(k)$, that its
actual degree is $k$, is: $P_i(k) = \lambda _i^k e^{-\lambda _i}/k! $ Hence, the overall degree distribution
$P(k) = \protect {\int _0^\infty } \protect \frac  {\lambda ^k}{k!} e^{-\lambda } \rho _\lambda (\lambda ) \protect \mathrm 
{d}\lambda $. We call this the \emph{intrinsic Poisson behavior}.
Although the effect of this behavior on the resulting
actual degree distribution persists for slowly decaying $\rho _r$, the asymptotic behaviors coincide \cite{Sokolov2010}.

Now, in the case of heavy-tailed distributions the random variable $r$ (which represents the $r_i$) can be expressed as a function
of a random variable $R$:
\begin{align}
   r = \frac{1}{1 + xR},
   \label{equ:r_is_1/1+xR}
\end{align}
where $R$ is a quotient of two independent random variables $\xi_1$ and $\xi_2$ distributed according to
the one-sided L\'evy stable pdf 
$L_\alpha$ \cite{Sokolov2010, Eliazar2005},
with $x = (M-1)^{1/\alpha} \approx M^{1/\alpha}$.
For such variables it is known that $\ew{\exp(-u\xi)}_\xi = \exp (-u^\alpha)$ and thus we can write:
\begin{align*}
  \ew{\exp(-uR)}_R &= \ew{\exp\left( -u \frac{\xi_1}{\xi_2} \right)}_{\xi_1, \xi_2}\\
  &= \ew{\exp\left[ -\left(\frac{u}{\xi_2}\right)^\alpha \right]}_{\xi_2}\\
  &= \int_0^\infty \exp\left[ -\left(\frac{u}{\xi_2}\right)^\alpha\right] L_\alpha \left(\xi_2 \right) \d \xi_2.
\end{align*}
This can be transformed 
by using the results of Chap. XIV in \cite{Feller2} to the Mittag-Leffler-Function
$E_\alpha(z) = \sum_{n=0}^\infty z^n/\Gamma(\alpha n +1)$:
\begin{align}
  \ew{\exp(-uR)}_R &= E_\alpha(-u^\alpha).\label{equ:LaplaceOfRhoR}
\end{align}
The details of the derivation of Eq. (\ref{equ:LaplaceOfRhoR}) are given in the appendices of \cite{Sokolov2010} and \cite{Eliazar2005}.
With this relation the asymptotic behavior of $\rho_R$ for large values of $R$ can be obtained via Corollary 8.1.7
of \cite{Bingham1987} which is an extension of the Tauberian theorems: From
\begin{align}
  1 - E_\alpha (-u^\alpha) \simeq \frac{u^\alpha}{\Gamma(\alpha + 1)} \quad \text{for} \quad &u \rightarrow 0\nonumber
  \intertext{it follows that}
   \rho_R(R) \simeq \frac{R^{-1-\alpha}}{\Gamma(\alpha)\Gamma(1 - \alpha)} \quad \text{for} \quad &R \rightarrow \infty.
   \label{equ:BehaviorOfRatInfty}
\end{align}
Because $R$ is a quotient of two identically distributed non-negative random variables, the distributions of $R$ and $1/R$ are the same.
Since
\begin{align}
   \frac{R^{-1-\alpha}}{\Gamma(\alpha)\Gamma(1 - \alpha)} \stackrel{R\rightarrow \infty}\simeq \rho_R(R) = \rho_{1/R} (R)
   = \frac{1}{R^2}\rho_R \left(\frac{1}{R}\right), \nonumber
\end{align}
we have
\begin{align}
   \rho_R\left( \frac{1}{R} \right) \simeq \frac{R^{1-\alpha}}{\Gamma(\alpha)\Gamma(1 - \alpha)} \quad \text{for} \quad R\rightarrow \infty.\nonumber
\end{align}
Thus, we can also determine the behavior of $\rho_R(R)$ for small $R$:
\begin{align}
   \rho_R\left( R \right) \simeq \frac{R^{\alpha-1}}{\Gamma(\alpha)\Gamma(1 - \alpha)} \quad \text{for} \quad R\rightarrow 0.
   \label{equ:BehaviorOfRatZero}
\end{align}
Using the connection between $\rho_r$ and $\rho_R$:
\begin{align}
  \rho_r(r) = \frac{1}{xr^2}\rho_R\left( \frac{\frac{1}{r} -1}{x} \right),
\end{align}
we can estimate the behavior of $\rho_r$ for small and large values of $r$.

For $r \rightarrow 0$ with $1/r \rightarrow \infty$, we can use Eq. (\ref{equ:BehaviorOfRatInfty}):
\begin{align}
   \rho_r(r) \simeq \frac{1}{xr^2}\frac{(rx)^{1+\alpha}}{\Gamma(\alpha)\Gamma(1 - \alpha)}
   &\quad \text{for}\quad r\rightarrow 0; 
\intertext{for $r \rightarrow 1$, i.e. $(1/r) - 1 \rightarrow 0$, we use Eq. (\ref{equ:BehaviorOfRatZero}):}
   \rho_r(r) \simeq \frac{r^{-1-\alpha}(1-r)^{\alpha-1}}{M\Gamma(\alpha)\Gamma(1 - \alpha)} &\quad \text{for}\quad r\rightarrow 1.
\end{align}
For $r \ll 1$ this behaves as $\rho_r(r) \simeq r^{-1-\alpha}/(M\Gamma(\alpha)\Gamma(1 - \alpha))$.

On the total $\rho_r$ shows the following regimes:

\newsavebox{\mycases}
\begin{subequations}
\begin{align}
  \sbox{\mycases}{$\displaystyle \rho_r(r) \simeq\left\{\begin{array}{@{}c@{}}%
  \vphantom{\frac{Mr^{\alpha-1}}{\Gamma(\alpha)\Gamma(1 - \alpha)} \text{for } r < M^{-1/\alpha},\pb{2.6em}}\\%
  \vphantom{\frac{r^{-1-\alpha}}{(M\Gamma(\alpha)\Gamma(1 - \alpha))} \text{for } M^{-1/\alpha} < r \ll 1,\pb{2.6em}}\\%
  \vphantom{\frac{r^{-1-\alpha}(1-r)^{\alpha-1}}{M\Gamma(\alpha)\Gamma(1 - \alpha)} \text{else.}\pb{2.6em}}\end{array}\right.\kern-\nulldelimiterspace$}
  \raisebox{-.7\ht\mycases}[0pt][0pt]{\usebox{\mycases}}\hphantom{\rho_r(r)}%
     {\textstyle \frac{Mr^{\alpha-1}}{ \Gamma(\alpha)\Gamma(1 - \alpha)}}  &\quad\text{for } r < M^{-1/\alpha},\pb{2.5em}\label{equ:rho_rCasesA} \\
     {\textstyle \frac{r^{-1-\alpha}}{M\Gamma(\alpha)\Gamma(1 - \alpha)}} &\quad\text{for } M^{-1/\alpha} < r \ll 1,\pb{2.5em}\label{equ:rho_rCasesB}\\
     {\textstyle  \frac{r^{-1-\alpha}(1-r)^{\alpha-1}}{M\Gamma(\alpha)\Gamma(1 - \alpha)}} &\quad\text{else.}\pb{2.5em}\label{equ:rho_rCasesC}
\end{align}
\end{subequations}
The transition between the first two regimes takes place at $r^\star = 1/x = M^{-1/\alpha}$.
The pdf $\rho_\lambda(\lambda)$ of degree expectation values follows:
\begin{align*}
   \rho_\lambda(\lambda) = \frac{1}{2N}\rho_r\left(\frac{\lambda}{2N}\right).
\end{align*}

\section{Simple Graphs}
Now we use this model to generate simple graphs.
One possibility to achieve this is to start with the ordinary
canonical fitness model and merge the multiple edges to a single one as it is done in \cite{Goh2001}.
The probability $p_{ij}$ that there is an edge between any two vertices $i$ and $j$ is due to Poissonian statistics:
\begin{align}
   p_{ij}	= 1 - e^{-2Nr_i r_j}\label{equ:linkingFct}.
\end{align}
Note that Eq. (\ref{equ:linkingFct}) can also be used to define the linking probability $F(f_i, f_j) = p_{ij}$ in the grand canonical model.

Merging the edges changes the degree distribution of a graph.
We first examine the degree expectation values $\nu_j = \langle k^{(s)}_j \rangle$ of vertices $j$ of this simple graph.

\subsection{Relation between Fitness and Degree Expectation}
In a network of $M$ vertices and $N$ thrown edges,
the expected degree $\nu_j$ of a node $j$ with normalized fitness
$r_j$ is:
\begin{align}
  \nu_j	&= \sum_{i=1}^M \left( 1-e^{-2N r_j r_i} \right)
		\cong M\left( 1- \int_0^1 e^{-2N r_j r} \rho_r(r)\d r\right).\nonumber
\intertext{Let $\lambda_j$ again denote $2Nr_j$ and consider $\nu$ as a function of $\lambda$:}
  \frac{\nu}{M}	&= 1- \int_0^1 e^{-\lambda r} \rho_r(r)\d r  = M\left( 1 - \left\langle e^{-\lambda r}\right\rangle \right). \label{equ:ExpExp}
\end{align}

Here the two cases $\lambda \ll 1$ and $\lambda \gg 1$ have to be considered separately.

For $\lambda \ll 1$, the argument of the exponential $\exp[-\lambda r]$ is small. Expanding the exponential into a Taylor series and
performing the integration leads in the lowest non-vanishing order to:
\begin{align}
  \frac{\nu}{M}	&= 1- \sum_{n=0}^\infty\frac{\left( -\lambda\right)^n}{n!}\langle r^n \rangle\label{equ:MomentExpansion}\\
				&\approx 1- \left( 1 - \lambda \langle r \rangle \right).\label{equ:MomentExpansion2}
\end{align}
The first moment of $r$ is given by:
\begin{align}
  \langle r \rangle = \int_0^\infty r(R) \rho_R(R) \d R = \int_0^\infty \frac{1}{1 + xR} \rho_R(R) \d R\label{equ:12}.
\end{align}
It was evaluated in \cite{Sokolov2010} by using:
\begin{align}
   \frac{1}{1 + xR} &= \frac{1}{x} \int_0^\infty e^{u/x}e^{-uR}\d u.\label{equ:UsefulIntegral}
\end{align}
For the sake of completeness we will recall the calculation here:
Substituting Eq. (\ref{equ:UsefulIntegral})  into Eq. (\ref{equ:12}) and changing the order of integration yields:
\begin{align*}
  \langle r \rangle &= \frac{1}{x} \int_0^\infty e^{u/x} \int_0^\infty e^{-uR} \rho_R(R) \d R \d u\nonumber\\
  &=  \frac{1}{x} \int_0^\infty e^{u/x} E_\alpha (-u^\alpha) \d u.\nonumber
\end{align*}
The last integral is the Laplace transform of the Mittag-Leffler function which is given by
$\mathcal{L}\{ E_\alpha (-u^\alpha) \} = s^{\alpha-1}/(s^\alpha +1)$ \cite{Feller2}. Setting $s = 1/x$, we find:
\begin{align}
   \langle r \rangle = \frac{1}{x^\alpha + 1} = \frac{1}{M}.
   \label{equ:FirstMomentOfr}
\end{align}
Thus, for $\lambda \ll 1$, we have found out that $\nu(\lambda) = \lambda$, i.e.
\begin{align}
  \nu(r)	&\propto 2Nr \qquad (\lambda \ll 1).\label{equ:NuSmall}
\end{align}

\begin{figure}
       \includegraphics[width=\linewidth]{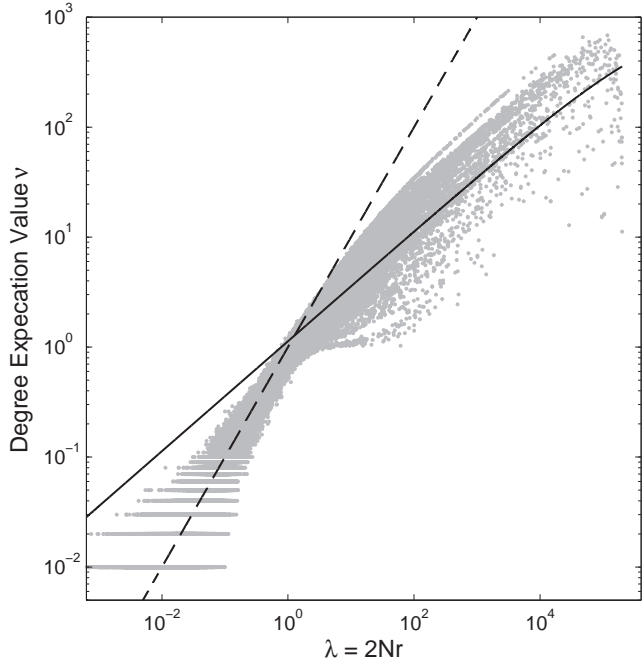}
       \caption{\footnotesize 
		      Expected degree $\nu$ of a
		      vertex in dependence on its normalized fitness parameter $r$.
		      The order of the networks is $M = 10^3$ and the size $N = 10^5$,
		      the L\'evy-exponent is $\alpha = 0.5$. The simulated data are shown by gray dots.
		      The dashed line shows the approximation of Eq. (\ref{equ:NuOfRA}): $\nu(\lambda) = \lambda$,
		      which holds for small $\lambda$.
		      For large values of $\lambda$, $\nu(\lambda)$ behaves as the solid line indicates, see Eq. (\ref{equ:NuOfRC}).
		      } 
       \label{fig:MeankMicroMerged}
\end{figure}

For $\lambda \gg 1$,
the expansion of the exponential into a power series as in Eq. (\ref{equ:MomentExpansion}) is not eligible
since the series in Eq. (\ref{equ:MomentExpansion}) converges slowly if at all.
Hence, for the calculation of $\left\langle \exp\left( -\lambda r\right)\right\rangle$, we resort to directly substituting Eq. (\ref{equ:r_is_1/1+xR})
into Eq. (\ref{equ:ExpExp}):
\begin{align}
  \frac{\nu}{M}	&= 1- \int_0^1 e^{-\lambda r} \rho_r(r)\d r
				= 1- \int_0^\infty e^{-\frac{\lambda}{1 + xR}} \rho_R(R)\d R.\label{equ:nudurchMistgleich}
\end{align}
Now we remember that $\omega = 1/R$ has the same pdf as $R$, and we write Eq. (\ref{equ:nudurchMistgleich}) as:
\begin{align}
  \frac{\nu}{M}	&= 1- \int_0^\infty e^{-\frac{\lambda \omega}{x + \omega}} \rho_R(\omega)\d \omega.\label{equ:FirstinOmega}
\end{align}
For large values of $\lambda$ the behavior of $\nu(\lambda)$ is determined by the
behavior of the integrand at small values of $\omega$, where the Eq. (\ref{equ:FirstinOmega}) can be approximated by:
\begin{align}
  \frac{\nu}{M}	&\approx 1- \int_0^\infty e^{-\frac{\lambda \omega}{x}} \rho_R(\omega)\d \omega\label{equ:SecondinOmega}.
\end{align}
The error of this approximation is of the order of $1/\lambda$.
Evaluating the integral in Eq. (\ref{equ:SecondinOmega}) and using Eq. (\ref{equ:LaplaceOfRhoR}) we get:
\begin{align}
  \frac{\nu}{M}	&= 1-E_\alpha\left[-\left(  \frac{\lambda}{x} \right)^\alpha\right]+ \mathcal{O}\left( \frac{1}{\lambda} \right)\\
				&\approx 1-E_\alpha\left(-\frac{\lambda^\alpha}{M}\right).\label{equ:NuLarge}
\end{align}
The asymptotes of Eq. (\ref{equ:NuSmall}) and Eq. (\ref{equ:NuLarge}) intersect at
$\lambda^\star \approx \left[ \Gamma(\alpha + 1) \right]^{1/(\alpha -1)}$.

Overall, we find:
\newsavebox{\mycasess}
\begin{subequations}
\begin{align}
  \sbox{\mycasess}{$\displaystyle \nu \simeq \left\{\begin{array}{@{}c@{}}%
  \vphantom{2Nr\pb{2.7em}}\\%
  \vphantom{M\left( 1 - E_\alpha\left( -\frac{(2Nr)^\alpha}{M} \right)\right)\pb{2.7em}}\end{array}\right.\kern-\nulldelimiterspace$}
  \raisebox{-.5\ht\mycasess}[0pt][0pt]{\usebox{\mycasess}}\hphantom{\nu}&%
     {\textstyle 2Nr} &&\qquad&\text{for } r \ll \frac{\lambda^\star}{2N},\pb{2.5em}\label{equ:NuOfRA} \\*
     &\rlap{${\textstyle M\left( 1 - E_\alpha\left( -\frac{(2Nr)^\alpha}{M} \right)\right)}$} &&\qquad&\text{else.}\pb{2.5em}\label{equ:NuOfRC}
\end{align}
\end{subequations}
%  The simulated data (gray dots) are generated by averaging
%  each of the 100 fitness realizations over 100 network realizations.
In Fig. \ref{fig:MeankMicroMerged} this is compared to simulation results obtained by studying networks of order
$M = 10^3$ and size $N = 10^5$,
with the L\'evy-exponent $\alpha = 1/2$. In the simulations we first generate a realization of the fitness distribution
and calculate the mean degrees of the vertices by averaging over 100 realizations of
the network. This is repeated for 100 fitness realizations following from the same fitness distribution.
For small values of $\lambda$ the averaged degrees behave as $\nu(\lambda) = \lambda$.

For $1 \ll \lambda \ll x$ we use the expansion of $E_\alpha$ for small values of the argument
$\nu/M	\simeq \frac{\lambda^\alpha}{M\Gamma(\alpha + 1)}$, thus:
\begin{align}
 \nu \simeq \frac{(2Nr)^\alpha}{\Gamma(\alpha + 1)} \qquad\text{for}\quad \frac{\lambda^\star}{2N} \ll r \ll \frac{M^{1/\alpha}}{2N}.\label{equ:NuOfRB}
\end{align}

\begin{figure}[t]
       \includegraphics[width=\linewidth]{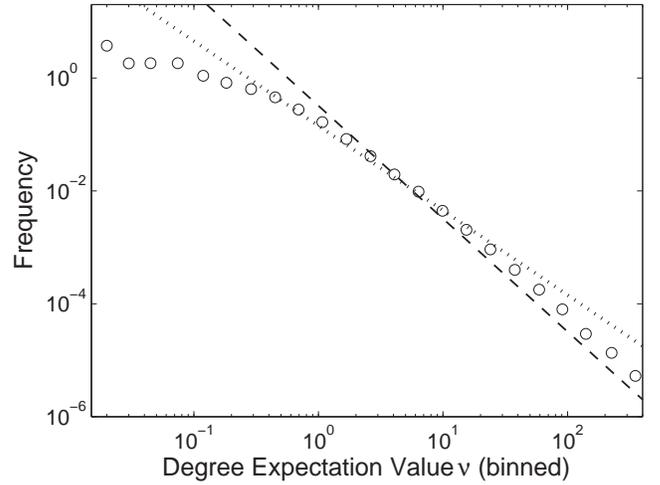}
       \caption{\footnotesize Distribution of the averaged degrees $\nu$ for $\alpha = 0.5$.
		      The order of the networks is $M = 10^3$ and the size $N = 10^5$.
		      The empty circles are the simulation results obtained by binning the data.
		      The dashed line shows the asymptotic $\nu^{-2}$-behavior. The dotted line represents the intermediate
		      asymptotic behavior $\rho_\nu(\nu) \propto \nu^{-1-\alpha}$.
		      }
       \label{fig:MeankMicroMergedDistr}
\end{figure}

\subsection{Distribution of Degree Expectations}
Using the dependence $\nu (r)$ as given by Eqs. (\ref{equ:NuOfRA}) and (\ref{equ:NuOfRC}), we can calculate the distribution $\rho_\nu (\nu)$ of
degree expectations in the considered network:
\begin{align}
   \rho_\nu (\nu) = \left| \frac{\d r (\nu)}{\d \nu} \right| \rho_r \left( r (\nu) \right) .
   \label{equ:Distr}
\end{align}

Eqs. (\ref{equ:NuOfRA}) and (\ref{equ:NuOfRC}) yield:
\newsavebox{\mycasesss}
\begin{subequations}
\begin{align}
  \sbox{\mycasesss}{$\displaystyle r \simeq \left\{\begin{array}{@{}c@{}}%
  \vphantom{2Nr\pb{2.7em}}\\%
  \vphantom{\frac{(2Nr)^\alpha}{\Gamma(\alpha + 1)}\quad\text{for } \frac{\lambda^\star}{2N} \ll r \ll \frac{M^{1/\alpha}}{2N},\pb{2.7em}}\\%
  \vphantom{M\left( 1 - E_\alpha\left( -\frac{(2Nr)^\alpha}{M} \right)\right)\pb{2.7em}}\end{array}\right.\kern-\nulldelimiterspace$}
  \raisebox{-.6\ht\mycasesss}[0pt][0pt]{\usebox{\mycasesss}}\hphantom{}&%
     {\textstyle \frac{\nu}{2N}} &&\hspace{-0.2cm}\text{for } \nu \ll \lambda^\star,\pb{2.5em}\label{equ:ROfNuA} \\*
     &{\textstyle \frac{\left( \nu \Gamma(\alpha +1)\right)^{1/\alpha}}{2N}} &&\hspace{-0.2cm}\text{for } {\textstyle \lambda^\star \ll \nu \ll \frac{M}{\Gamma(\alpha + 1)}},\pb{2.5em}\label{equ:ROfNuB}\\*
     &{\textstyle \frac{\left( -ME_\alpha^{-1} \left( 1 - \frac{\nu}{M} \right) \right)^{1/\alpha}}{2N}} && ~\text{else,}\pb{2.5em}\label{equ:ROfNuC}
\end{align}
\end{subequations}
where we have singled out explicitly the intermediate regime Eq. (\ref{equ:ROfNuB}).
Together with Eqs. (\ref{equ:rho_rCasesA}) to (\ref{equ:rho_rCasesC}), this gives the distribution of degree expectation values $\rho_\nu$.
We assume that the considered
networks are sparse, i.e. $p = 2N/M^2 \ll 1$ and hence the condition of
Eq. (\ref{equ:NuOfRC}) or Eq. (\ref{equ:ROfNuC}), respectively, does not apply.
Moreover, we concentrate on the values of $r$ in $M^{-1/\alpha} < r \ll 1$ where
$\rho_r(r) \sim r^{-1-\alpha}$.
The smaller values of $r$ are irrelevant for networks of moderate size and the large values of $r$ only
describe the cut-off of the actual power-law.
Thus, we have to examine the two cases:
\begin{align}
   \rho_\nu(\nu) \simeq \frac{1}{2N}\rho_r\left( \frac{\nu}{2N}\right)
\end{align}
for the domain of Eq. (\ref{equ:ROfNuA}) and
\begin{align}
   \rho_\nu(\nu) \simeq \frac{\left[\Gamma(\alpha +1)\right]^{1/\alpha} \nu^{1/\alpha - 1}}{2N\alpha} \rho_r\left( \frac{\left[ \nu \Gamma(\alpha +1) \right]^{1/\alpha}}{2N} \right)
\end{align}
for the domain of Eq. (\ref{equ:ROfNuB}).
This gives two asymptotes which intersect at $\nu^\star = \left[ \alpha \Gamma(\alpha + 1) \right]^{1/(\alpha -1)}$:
\newsavebox{\mycasessss}
\begin{subequations}
\begin{align}
  \sbox{\mycasessss}{$\displaystyle \rho_\nu(\nu) \simeq \left\{\begin{array}{@{}c@{}}%
  \vphantom{2Nr\pb{2.7em}}\\%
  \vphantom{M\left( 1 - E_\alpha\left( -\frac{(2Nr)^\alpha}{M} \right)\right)\pb{2.7em}}\end{array}\right.\kern-\nulldelimiterspace$}
  \raisebox{-.5\ht\mycasessss}[0pt][0pt]{\usebox{\mycasessss}}\hphantom{}&%
     {\textstyle \frac{\nu^{-1-\alpha} (2N)^\alpha}{M\Gamma(\alpha)\Gamma(1-\alpha)}} &&\hspace{-0.2cm}\text{for } {\textstyle \frac{2N}{M^{1/\alpha}}} < \nu < \nu^\star,\pb{2.5em}\label{equ:rhoNuA} \\*
     &{\textstyle \frac{\nu^{-2} (2N)^\alpha}{M\left[\Gamma(\alpha +1)\right]^2\Gamma(1-\alpha)}} &&\hspace{-0.2cm} \text{for } \nu^\star < \nu.\pb{2.5em}\label{equ:rhoNuB}
\end{align}
\end{subequations}
In Fig. \ref{fig:MeankMicroMergedDistr} this is compared to simulation results with networks of the order $M = 10^3$, the size $N = 10^5$,
and the L\'evy-exponent $\alpha = 1/2$.
Eq. (\ref{equ:rhoNuB}) implies that
for large degree expectation values $\nu$ the pdf $\rho_\nu(\nu)$ follows
the power-law: $\rho_\nu(\nu) \sim \nu^{-2}$ regardless of the exact value of $\alpha$. This $\nu^{-2}$-behavior can be
seen in Fig. \ref{fig:MeankMicroMergedDistr}.
For $\alpha \leq 1/2$ the first regime does always exist because then
$2Nr^\star<2N/M^2=p<1< \left[ \alpha \Gamma(\alpha + 1) \right]^{1/(\alpha -1)}$
(see Fig \ref{fig:MeankMicroMergedDistr}).
For $\alpha > 1/2$, the existence of the first regime depends on how sparse the network is.
If $\left[ \alpha \Gamma(\alpha + 1) \right]^{1/(\alpha -1)} < 2N/M^{1/\alpha}$,
this regime is not visible, and  the $\nu^{-2}$-behavior
is observed over the whole domain of relevant $\nu$. This gives another example of
a universal inverse square distribution \cite{DeLosRios2001,Caldarelli2004,Goh2002}.
The asymptotics of degree distributions $P(k)$ follows the asymptotics of $\rho_\nu (\nu)$.

\section{Summary}
Many real world networks exhibit degree distributions which follows a power-law $P(k) \propto k^{-1-\alpha}$.
The generalizations of classical models including intrinsic fitness parameters are considered to be applicable to
generate such networks.

We discussed in detail one of such models concentrating especially on the domain of $0<\alpha<1$ in which the corresponding
distributions do not posses a mean. We first examined the situations in which the model is used to produce a multigraph.
Then we studied what happens if the multiple edges are merged to a single one and thus a simple graph is built.
We gave the relation between the
(normalized) fitness parameter $r$ and the expected degree $\nu$ of a vertex 
and showed that it possesses non-trivial intermediate and final asymptotic behaviors. Especially interesting is the fact that for large values
of expected degrees $\nu$ the final asymptotics is universally $\rho_\nu(\nu) \propto \nu^{-2}$ and does not depend on $\alpha$.

\section{Acknowledgements}
The financial support of DFG within the IRTG 1740 ``Dynamical Phenomena in Complex Networks: Fundamentals and Applications'' is gratefully acknowledged.

\bibliographystyle{ieeetr}

\end{document}